\documentclass[letterpaper]{JHEP3}
\preprint{}
\keywords{M-Theory, Anomalies in Field and String Theories}

\usepackage{amscd,verbatim}
\usepackage[all]{xy}
\usepackage{graphicx}

\usepackage{amsmath}

\newcommand{\bp}{\mathop{\vtop{\ialign{##\crcr

$\hfil\displaystyle{}\hfil$\crcr\noalign{\kern-13pt\nointerlineskip}
       \BIG{(}\hskip0pt\crcr\noalign{\kern3pt}}}}}
\newcommand{\cbp}{\mathop{\vtop{\ialign{##\crcr

$\hfil\displaystyle{}\hfil$\crcr\noalign{\kern-13pt\nointerlineskip}
       \BIG{)}\hskip0pt\crcr\noalign{\kern3pt}}}}}
\newcommand{\pa}{\mathop{\vtop{\ialign{##\crcr

$\hfil\displaystyle{\oplus}\hfil$\crcr\noalign{\kern+1pt\nointerlineskip
}
       \hspace{.08in}$^{\alpha=0}$\hskip6pt\crcr\noalign{\kern3pt}}}}}

\newcommand{\p}{^\prime}

\newcommand{\rank}{{\textup{\scriptsize{rank}}}}

\newcommand{\R}{\ensuremath{\mathbb R}}

\newcommand{\cL}{\ensuremath{\mathcal L}}

\newcommand{\cT}{\ensuremath{\mathcal T}}

\newcommand{\Z}{\ensuremath{\mathbb Z}}

\def\dwn{\downarrow}

\def\L{\ensuremath{{\cal L}}}

\def\O{\ensuremath{{\cal O}}}

\newcommand{\beq}{\begin{equation}}
\newcommand{\eeq}{\end{equation}}

\newcommand{\tensor}{\otimes}

\numberwithin{equation}{section}

\def\hsp#1{\hspace{#1in}}

\catcode`\@=11
\def\vereq#1#2{\lower3pt\vbox{\baselineskip1.5pt \lineskip1.5pt
\ialign{$\m@th#1\hfill##\hfil$\crcr#2\crcr\sim\crcr}}}
\catcode`\@=12

\makeatletter
\newcommand\figcaption{\def\@captype{figure}\caption}
\newcommand\tabcaption{\def\@captype{table}\caption}
\makeatother
\renewcommand{\(}{\begin{equation}}
\renewcommand{\)}{\end{equation}}

\newcommand{\RR}{{\mathbb R}}
\newcommand{\ZZ}{{\mathbb Z}}

\newcommand{\attn}[1]{\begin{center}\framebox{\begin{minipage}{12cm}#1
\end{minipage}}\end{center}}

\title{Some relations between twisted K-theory and $E\sb8$ gauge
theory}

\author{Varghese Mathai\thanks{VM is financially supported by the Australian Research Council}\\
Department of Pure Mathematics \\
University of Adelaide \\
Adelaide, SA 5005 \\
Australia\\
Email: \email{vmathai@maths.adelaide.edu.au}}

\author{Hisham Sati\thanks{HS is financially supported by the Australian
Research Council}\\
Department of Physics and Mathematical Physics, and \\
Department of Pure Mathematics \\
University of Adelaide \\
Adelaide, SA 5005 \\
Australia\\
Email: \email{hsati@maths.adelaide.edu.au}}

\abstract{
Recently, Diaconescu, Moore and Witten provided a nontrivial link
between K-theory and M-theory, by deriving the partition function of the
Ramond-Ramond fields of Type IIA string theory from an $E_8$ gauge 
theory
in eleven dimensions.
We give some relations between twisted K-theory and
M-theory by adapting the method of \cite{DMW1}, \cite{MS}. In particular,
we construct the twisted K-theory torus which defines the partition
function, and also discuss the problem from the loop group picture,
in which the Dixmier-Douady class is the Neveu-Schwarz field.
In the process
of doing this, we encounter some  
mathematics that is new to the physics literature.  In  particular, the eta
differential form, which is the generalization of the eta invariant,
arises naturally in this context.
We conclude with several open problems in mathematics and string theory.}

\begin{document}

\section{Introduction and setup}
Type II string theories in ten dimensions contain in addition to 
gravity
and fermions,
p-form fields, the Ramond-Ramond RR and the Neveu-Schwarz NS fields.
D-branes are charged \cite{Pol} under those p-forms.
It is by now well known that RR charges in the absence of NS fields
can be classified by  K-theory of spacetime \cite{MM}, namely by
$K^0(X)$
for type IIB \cite{Wi1} and by $K^1(X)$ for type IIA \cite{Ho}.
The RR fields are also classified by  K-theory \cite{MW, FH}, with the
roles of $K^0$ and $K^1$ interchanged. In the presence of a NS 
B-field, or its field strength $H_3$, the fields and the charges are
classified by
twisted K-theory, in the sense of \cite{Ros}, as was shown in 
\cite{FW, Ka}
by analysis 
of worldsheet anomalies for the case the NS field $[H_3] \in H^3(X,
\mathbb Z)$ 
is a torsion class, and in
\cite{BM} for the nontorsion case.

M-theory is a theory in eleven dimensions which is not yet known except
in specific regions (or points) of its "moduli space". It has been shown 
by
Witten \cite{Wi2, Wi3} that the topological part can be encoded in the
index theory of an $E_8$ gauge bundle.
At the level of supergravity, the low energy limit of string theories 
and
M-theory, there is an explicit relation between the two given by
Kaluza-Klein reduction. The story is much more subtle at the quantum 
level
due to the existence of nontrivial phase factors in the partition
functions. In a rather nontrivial way, it has been recently shown by
\cite{DMW1, DMW2} that one can
also relate the corresponding partition functions, namely the one 
derived
using the $E_8$ theory in eleven dimensions and the one derived from
K-theory in ten dimensions. The authors restricted themselves mostly
to the RR sector. This has been generalized in \cite{MS}
to include the fermions, one-loop contributions and membrane instantons,
as well as including flat background NS potentials. The authors show,
nontrivially, that the partition functions are
T-duality invariant only after including the above effects. They also
identify the T-duality anomalies. In both \cite{DMW1, DMW2} and
\cite{MS} (see also \cite{M}), it has been suggested to include nontrivial
$[H_3]$.

In this paper, we attempt at generalizing some of the ideas along the 
lines
of \cite{DMW1, DMW2, MS} in the context of twisted K-theory.
A convenient computational tool for twisted K-theory is the
Atiyah-Hirzebruch spectral sequence (AHSS) \cite{AHSS}, which has been
nicely used 
in
analyzing physical D-brane configurations (see e.g. \cite{MMS1, MMS2,
J2}). Inspite of the apparent physical favor of AHSS, we prefer here to
work in the full twisted K-theory.

The spaces we deal with are the following.
$Y$ is an eleven-dimensional spin manifold corresponding to M-theory.
$X$ is a ten-dimensional manifold which is the base for a circle
bundle with total space $Y$, and corresponds to Type IIA superstring
theory. Finally, $Z$ is a twelve-dimensional manifold which is a disk
bundle over $X$, whose boundary is the circle bundle $Y$
over $X$.

The basic setup for the bundles we consider is given in the following
diagram
\(
\begin{matrix}
E_8&\to & P\cr
&&\dwn\cr
S^1& \to & Y\cr
&&\dwn \pi\cr
     & & X\cr
\end{matrix}
\)
where $P$ is a principal $E_8$ bundle over the $11$-dimensional manifold
$Y$, which in turn is a principal $S^1$ bundle over the
$10$-dimensional manifold $X$.  Then $Y$
has a supergravity field whose field strength is a closed
$4$-form $G_4$,  that is related
to the integral characteristic class invariant $a$ of  $P$ as follows:
\begin{equation}
     \frac{G_4}{2\pi} = a - \frac{\lambda}{2}
\end{equation}
where $ \frac{\lambda}{2}$ is equal to half the first Pontrjagin class
$p_1(Y)$ of $Y$.
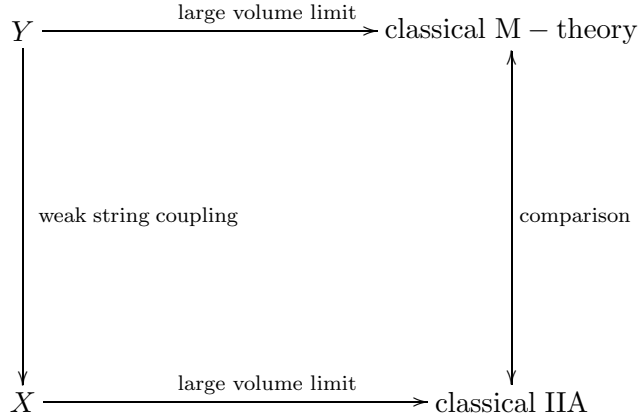
\bigskip
\begin{equation} \label{correspondence}
\xymatrix @=5pc {
Y   \ar@{->}[rr]^{\textup{large volume limit}} &&  {\rm classical
\;M-theory}\\
&& \\
X \ar@{<-}[uu]_{\textup{\ weak\;string\;coupling}}
\ar@{->}[rr]^{\textup{large volume limit}}  && {\rm classical\; IIA}
\ar@{<->}[uu]_{\textup{comparison}}
}
\nonumber
\end{equation}

\bigskip

We are interested in the comparison, using the metric
     $g_{Y} =  t\pi^*(g_{X})
+  \pi^*(e^{2 \phi/3}) {\cal{A}}\otimes  {\cal{A}} $,
in the large volume, adiabatic limit as $t \rightarrow \infty$. We explain 
the notation in section \ref{Rdxn}.

This paper is organized as follows. In section \ref{M} we review the
relation between M-theory and $E_8$ gauge theory, and then the
M-theory
partition function in section \ref{MPF}, and write the phase in terms
of the reduced eta invariant. In section \ref{Rdxn} we relate the eleven
dimensional fields to the ten dimensional ones by dimensional reduction,
and relate the eta invariant in the adiabatic limit to an integral
involving the eta form. In section \ref{IIA} we study the twisted K-theory
description of type IIA partition function and in particular the 
twisted
K-theory theta functions. We relate the $E_8$ bundle on $Y$ to an
$LE_8$ bundle on $X$, which gives rise to the
Neveu-Schwarz $H_3$ field (i.e. the twist), as well as a class in twisted
K-theory (over the rationals). In section \ref{dis} we conclude with
discussions and some open problems.

\section{M-theory}\label{M}
M-theory \cite{Wi0, T, Duff} has three kinds of impurities: membranes,
fivebranes and boundaries.
The low energy theory is eleven-dimensional supergravity. The massless
degrees of freedom are the metric $g$ , a three-form potential $C_3$,
and a Rarita-Schwinger fermionic spin $3/2$ field $\psi_M$.
The action of eleven dimensional supergravity is \cite{CJS}
\(\label{eleven}
I_{11}=I_{grav} + I_{G_4} + I_{C.S.}+ I_{fermi} + I_{coupling}
\)
where

\begin{eqnarray}
I_{grav} &= &\frac{1}{2 \kappa_{11}^2} \int_{Y}  {\hat{\cal{R}}} 
\;d {\rm vol} \\
I_{G_4} &= & - \frac{1}{2 \kappa_{11}^2} \frac{1}{2 \cdot 4!} \int_{Y}
| G_4 |^2\; d {\rm vol}\\
I_{C.S.} &=& -\frac{1}{12 \kappa_{11}^2} \int_{Y} C_3 \wedge G_4 \wedge
G_4
\\
I_{fermi} & =& \frac{1}{2 \kappa_{11}^2} \frac{1}{2} \int_{Y}
\bar{\psi}
D_{R.S.} \psi \; d {\rm vol}
\end{eqnarray}
where $d {\rm vol}= d^{11}x \sqrt{-g}$, $\hat{\cal{R}}$ is the scalar
curvature of $Y$, $G_4$ is the
four-form
field
strength, which, when cohomologically trivial, is equal to $dC_3$. The
fermions involve the kinetic action of $\psi_M$ involving
the Rarita-Schwinger operator $D_{R.S}$.  One can view $D_{R.S.}$
as \cite{DMW1, DMW2} the Dirac operator coupled to the vector bundle
associated to the
virtual bundle $TY - 3{\O}$, where the ${\O}$ factors correspond to
subtraction of ghosts.
$I_{coupling}$ corresponds to coupling of
$\psi_M$
to $G_4$ as well as quartic $\psi_M$ self-couplings. It is not essential
for our discussion and thus we do not record it.

The source-free Bianchi identity and equation of motion are
\begin{eqnarray}
dG_4&=&0
\\
d *G_4&=& -\frac{1}{2} G_4 \wedge G_4
\label{EOM}
\end{eqnarray}
which can be modified by adding sources, namely the membrane $M2$
and the fivebrane $M5$, respectively.
They have worldvolumes, respectively, ${\cal{W}}_3$ and ${\cal{W}}_6$
embedded
by
an embedding $\iota$ in spacetime $Y$. There are also one-loop 
corrections that, for example, modify the RHS of \eqref{EOM}
by the topological quantity $X_8=\frac{1}{192}(p_1^2 - 4p_2)$, given
in terms of the Pontrjagin classes of the tangent bundle $TY$.

The four-form of M-theory obeys the quantization condition \cite{Wi2}
\(
\frac{G_4}{2 \pi}+ w_4 \in H^4(Y ;\Z)
\)
where $w_4$ is the fourth Stiefel-Whitney class of the tangent
bundle
$TY$. In the orientable case, which is what we are interested in,
$w_4=\frac{\lambda}{2}$ mod 1, with $\lambda=\frac{p_1}{2}$.

\section{M-theory Partition Function} \label{MPF}

  This section is mostly a review of relevant facts. The
M-theory partition function is a product of factors corresponding 
to
the different parts of the action \eqref{eleven},

\(
Z_M \sim Z_{grav} Z_{G_4} Z_{C.S.} Z_{fermi} Z_{coupling}
\)

We are interested in the topological part of the partition function,
which
means that,as in \cite{DMW1}, we keep only the moduli associated with
$C_3$ (or $G_4$) but
keep
all the phases, so that the part we are interested in is
\(
e^{-||G_4(a)||^2} \Omega_M (C_3).
\)
Consequently, we do not consider $Z_{grav}$ nor $Z_{coupling}$.
This theory can be viewed as having two kinds of fermions
First, the
spin
$1/2$
fermions in the $E_8$ gauge theory
\footnote{note added in the proof: we do not address whether or not those
are physical fermions.
Some analysis on this (and on the question of
supersymmetry) can be found in \cite{inverse}.}
, and then the spin $3/2$
Rarita-Schwinger fields in the supergravity. At the level of actions, we
have
\(
I_M= I_{E_8} + \frac{I_{R.S.}}{2}
\)
The low energy quantum measure of M-theory factorizes in terms of
manifestly well-defined factors
\footnote{For the second factor $J=e^{iE_8}$: if $\partial Y \neq
\emptyset$ then \cite{Wi3} $J$ is not gauge-invariant, but is a section
of a line
bundle ${\L}^{-1}$ over the space of $C$-fields over $N=\partial Y$.}
\(
\det D_{R.S.}e^{iI_M} = \{\det D_{R.S.}e^{iI_{R.S.}/2}\} \cdot
e^{iI_{E_8}}
\)

The expression $G_4=dC_3$ is not valid globally and $C_3$ is not a
well-defined
differential form, implies that one has to be careful in defining the
topological part $I_{C.S.}$ of the action $I_{11}$.

The way around this is to lift to twelve dimensions and look at the
action

\(
I_{12} \sim \int_Z G_4 \wedge G_4 \wedge G_4.
\)
over a twelve dimensional manifold
$Z$. The full Chern-Simons coupling of M-theory is associated with
$I_{12}$, which is well-defined and independent of the choice of $Z$ and
of
the extension of $G_4$.
The action can be written as
\(
I_{12}=-\frac{1}{6} a^3
\)
where $a$ is the cohomology class of $\left[ \frac{G_4}{2 \pi} \right]$.
Witten \cite{Wi2, Wi3} have shown that there are two modifications to
this: First
that
$a - \frac{\lambda}{2}$ is
integral, and second we have to include $C_3 \wedge X_8$.
Introduce
an $E_8$ bundle $V$ on $Z$ whose characteristic class $\omega$ obeys
$ \omega=a - \frac{\lambda}{2}$. Witten have shown that
\(
\frac{I_{12}}{2 \pi}= \frac{i(E_8)}{2} + \frac{i(R.S.)}{4}
\)

and, including the above effects, the action takes the form

\(
\frac{I_{12}}{2 \pi}= -\frac{1}{6} \left(\omega -
\frac{\lambda}{2}\right)
\left[\left(\omega - \frac{\lambda}{2}\right)^2 -
\frac{1}{8}(p_2 - \lambda^2)\right]
\)
which is just $I_{C.S.}$ with the gravitational corrections turned on.

As for the Rarita-Schwinger path integral,
$\frac{1}{2 \pi}I_{12}$ can be half integral in general and has an
anomaly
that is cancelled from the one coming from the
determinant of the Rarita-Schwinger operator $\det D_{R.S.}$.
The combination shows up as

\(
\det D_{R.S.} e^{iI_{R.S.}/2}.
\)

The Rarita-Schwinger operator can be viewed as \cite{DMW1, DMW2}
the Dirac operator coupled to
$TX - 2{\O}$, since $Y$ is a circle bundle over $X$, or equivelently,
to $TZ - 4{\O}$ in twelve dimensions. Overall, one has the factor
\(
{\rm Pf}(D_{R.S.}) \exp{\left(i \int_Z I_{12}\right)}
\)
Now ${\rm Pf}(D_{R.S.})$ is a vector in a Pfaffian line, so the above 
can be
factorized into a modulus $|{\rm Pf} (D_{R.S.})|$ and a phase
$\Omega_M (C_3)\equiv (-1)^{I_{R.S.}/2} \exp{\left(i \int_Z
I_{12}\right)}$.

Using the Atiyah-Patodi-Singer (APS) index theorem \cite{APS1, APS2,
APS3} one can relate the
action to an index corrected by the reduced eta invariant
$\overline{\eta}=\frac{h+ \eta}{2}$, as

\(
I(D)=\int_Z i_D - \overline{\eta}
\)
so that the relevent integral in twelve dimensions can be written as

\(
\int_Z \frac{I_{12}}{2 \pi}=\frac{1}{2}I_{E_8} + \frac{1}{4}I_{R.S.}
+ \frac{h_{E_8} + \eta_{E_8}}{4} +  \frac{h_{R.S.} + \eta_{R.S.}}{8}
\)

Now the factor $(-1)^{I_{R.S.}}$ cancels the one coming from the index
theorem, and taking into account the fact that the index is even, the
phase derived in \cite{DMW1, DMW2} is
\(
\Omega_M (C_3)= \exp\left[2\pi i \left(
\frac{{\overline\eta}(D_{V(a)})}{2}
+ \frac{{\overline\eta}(D_{R.S.})}{4}\right)\right]
\label{phase}
\)

\section{Dimensional reduction from $Y$ to $X$}\label{Rdxn}

In \cite{DMW1, DMW2}, it was assumed that $C_3$ is a pullback from $X$.
This implies that the topological invariant $\Omega_M (C_3)$ depends 
only
on $a$ and {\it not} on $C_3$, i.e. $\Omega_M (a)$. We would like to 
study
the generalization of this to the case when the bundles in M-theory are
not lifted from the Type IIA base, and so we consider the case of
nontrivial Neveu-Schwarz H-field.

\subsection{Reduction of the Riemannian metric}

The Riemannian metric on
the circle
bundle $Y$  is $g_{Y} =  \pi^{\ast}(g_{X})
+  \pi^*(e^{2 \phi/3}) {\cal{A}}\otimes  {\cal{A}} $,
where $g_X$ is the Riemannian metric on $X$,
$e^{2 \phi/3}$ is the norm of the Killing vector along $S^1$, (which, in
this trivialization, is given by $\partial_z$) $\phi$ is the dilaton,
i.e. a real function on $X$ and  $ {\cal{A}}$ is a connection 1-form on
the circle
bundle  $Y$. Note that the component of the curvature in the direction
of the
circle action is
\(
R_{11}=e^{2\phi/3}=g_s^{2/3}.
\label{coupling}
\)
Such a choice of Riemannian metric is compatible with the principal
bundle
structure in the sense that the given circle action acts as isometries
on $Y$.

Performing a rescaling to the above metric and using the identification
\eqref{coupling}, the desired metric ansatz for IIA is
\(
g_Y= g_s^{4/3} g_{S^1} + t g_s^{-2/3} g_X
\)
in the limit $t \rightarrow \infty$ {\it{then}} $g_s \rightarrow 0$.

\subsection{Reduction of the differential forms $G_4$ and $G_7 = *G_4$}\label{twistform}
The reduction of the 4-form $G_4$ on $Y$ gives rise to two differential 
forms on $X$, the Neveu-Schwarz
3-form $H_3$ and the Ramond-Ramond 4-form $F_4$. This is obtained as
follows (setting the dilaton to a constant for simplicity).
For an oriented $S^1$ bundle
with first Chern class $c_1(Y)=F_2= d{\cal{A}}\in H^2(X,\ZZ)$,
we have
     a long exact sequence in cohomology
called the {\it Gysin sequence} (cf.~\cite[Prop.\ 14.33]{BT}).
\begin{equation}\label{gysin}
\begin{CD}
       \ldots @>>> H^k(X,\ZZ) @>\pi^*>> H^k(Y,\ZZ) @>\pi_*>>
H^{k-1}(X,\ZZ)
       @> F \cup >> H^{k+1}(X,\ZZ) @>>> \ldots
\end{CD}
\end{equation}
In  particular, with $k=4$, one sees that $ F_2 \cup \pi_* G_4 = dF_4$,
where $F_4$ is some differential 4-form on $X$. It follows that
$d( {\cal{A}}  \wedge \pi_* G_4 + F_4) = 0$. Therefore setting
$H_3 =  \pi_* G_4$, we see that $H_3$ is a closed form.  Noting that $\pi_*( {\cal{A}} ) = 1$,
we arrive at the equation on $Y$ (where it is understood that forms
on $X$ are pulled back to $Y$ via $\pi$)
\(\label{eqn:G_4}
G_4=F_4 + {\cal{A}} \wedge H_3.
\)
Now the curvature 2-form $F_2 = d {\cal{A}}$ is basic, i.e., it is
horizontal
$i_v F_2=0$, and invariant ${\cL}_v F_2=0$ where $v$ is a vertical vector. In a local trivialization of
the circle bundle where ${\cal{A}}=dz + \theta$ with $\theta$ being the
connection on $X$, the above two conditions mean, respectively,
that $F_2$ has no $dz$ component and that it does not depend explicitly
on $z$. 
Similarly, $G_4$ can be written in
the given trivialization, as
$G_4=F_4 + dz \wedge H_3$.


Suppose that $G_4 \in \Omega^4(Y)$ and the curvature $F_2\in \Omega^2(X)$ satisfies the Bianchi identities
 on $Y$ that are given below, and
which are obtained from the Euler-Lagrange equations for the Bosonic part of the 
action of eleven dimensional supergravity (cf. the formulae in equation \eqref{eleven}), namely 
$I_{grav} + I_{G_4}$,\footnote{These equations admit solutions for a particular ansatz. For 
example, when $X_8=0$ and $G_4$ is proportional to a volume form of a four-dimensional
factor in $Y$, this is the famous Freund-Rubin ansatz \cite{FR}. When 
$G_4$ is a flux through four-cycle(s) in $Y$, there are
solutions with $X_8\neq0$, cf. \cite{Bec}, for different choices of $Y.$}
\begin{eqnarray}
dG_4&=& 0,
\\
dG_7&=& -\frac{1}{2} G_4 \wedge G_4 + X_8. 
\label{eqn:Maxwell}
\end{eqnarray}
where $G_7 = *_{11}G_4$ and $X_8$ is a basic  differential form of degree $8$ on $Y,$
which is a Chern-Simons correction factor put in by hand.
 By applying the deRham differential
on both sides of equation \eqref{eqn:Maxwell}, we see that $X_8$ is a closed form.
\footnote{The one-loop coupling $\int C_3 \wedge X_8$
reduces to $\int B_2 \wedge X_8$ \cite{VW, DLM}.}

As argued above, the Bianchi identity $dG_4=0$ reduces to the Bianchi
identities for
the RR 4-form, the NS 3-form and the RR
2-form field
strengths,
respectively,
\begin{eqnarray}
dF_4&=& H_3 \wedge F_2,  
\\
dH_3&=&0,
\\
d{F}_2&=&0.
\end{eqnarray}

From general principles, we can write $\; G_7 = H_7 +{ \cal{A}} \wedge F_6$
where $H_7$ and $F_6$ are basic forms on $Y$ - this is consistent with equation \eqref{eqn:G_4}
since a standard computation shows that $i_v(*_{11}F_4) = *_{10}F_4$ and $i_v(*_{11}(A\wedge H_3))=0.$ 
Then, using equation \eqref{eqn:Maxwell},
one has
\begin{eqnarray}
dG_7 &=& dH_7  + F_2\wedge F_6 - {\cal{A}}\wedge dF_6,
\\
& =&-\frac{1}{2} F_4 \wedge F_4 -  {\cal{A}}\wedge H_3 \wedge F_4 + X_8,
\end{eqnarray}
Eliminating $dG_7$ from the equations above, one arrives at
\begin{equation}\label{eqn:preBianchi}
dH_7  =  - F_2\wedge F_6 + {\cal{A}}\wedge (dF_6   -H_3 \wedge F_4) - \frac{1}{2} F_4 \wedge F_4  + X_8.
\end{equation}
All of the terms in equation \eqref{eqn:preBianchi} are basic differential forms, with the sole exception of the 
term involving  $ {\cal{A}}$. Therefore contracting the terms of equation  \eqref{eqn:preBianchi} 
with the vertical vector field $v$, and using the fact that $i_v(\mathcal A) = 1$ and 
$i_v(dF_6   -H_3 \wedge F_4) = 0$,
we deduce  the corresponding ten-dimensional Bianchi identities on $X$,
\begin{eqnarray}
dF_8 & =& H_3 \wedge F_6,  \qquad dF_4= H_3 \wedge F_2,
\\
dF_6 & =&H_3 \wedge F_4, \qquad  dH_3 = 0, \quad dF_2=0,
\\
dH_7 & = &  - \frac{1}{2} F_4 \wedge F_4 - F_2\wedge F_6    + X_8,
\end{eqnarray}
where $F_6=*_{10}F_4$ and $F_8=*_{10}F_2$.

Summarizing our discussion, from $G_4 \in \Omega^4(Y)$ satisfying the eleven dimensional Bianchi identities, 
we obtain $F = F_2 + F_4 + F_6 + F_8 \in \Omega^{even}(X)$ satisfying
$(d-H_3 \wedge) F = 0$, where we observe 
that\footnote{hence note that, unlike the $[H_3]=0$ case \cite{MS}, we do not 
work in the massive Type IIA theory \cite{Rom}. }
$F_0 = 0$ since $H_3$ is not exact
and $F_8 \wedge H_3 =0$ for dimension reasons. Therefore 
$F$ determines a class in the twisted cohomology $H^{even}(X, H_3)$, 
where $H^{\bullet}(X, H_3)$ 
denotes the twisted cohomology, which is by definition
the cohomology of the  $\mathbb Z_2$-graded complex $(\Omega^\bullet(X), d - H_3 \wedge)$, where the de
Rham differential
is replaced by $d-H_3\wedge$, cf.  \cite{BCMMS} (see also \cite{FHT}).
Our discussion in this section can also be summarized by the following
diagram.
\begin{equation} \label{eqAc}
\xymatrix @=2pc { & G_4\in H^4(Y, \mathbb Z) + \text{(Bianchi)} \ar[dl]^{}
\ar[dr]_{} & \\
F \in H^{even}(X, H_3)  &   &
H_3\in H^3(X, \mathbb Z) }
\end{equation}


\subsection{Relating M-theory to K-theory}
We are dealing with Dirac operators coupled to certain vector bundles.
We are interested in the general case where the vector bundles
are not lifted from the base. First we have the twisting by the tangent
bundle, which leads to the Rarita-Schwinger operator. For this, one is
dealing with natural bundles and so are lifted from the base. However,
we also have the Dirac operator coupled to an $E_8$ vector bundle, which
we would like to consider as not lifted from $X$. This leads to the
appearance of eta-forms in the adiabatic limit of the reduced eta
invariant of that Dirac operator.

\subsubsection{Rarita-Schwinger Operator}
First we will look at $D_{R.S.}$. There are two contributions, one from
$h$ and the other from $\eta$.
Recall that in
dimensions $8n+2$, $h_{D\otimes V_{\RR}}$ is a topological invariant
mod 2. For the contribution from $h$, the idea is
to try to
relate the spectrum on $Y$ to that on $X$.
The authors of \cite{DMW1, DMW2} choose functions $\Phi$ that transform 
as
$\Phi \rightarrow
e^{-ik\theta}\Phi$ under an $S^1$-rotation by an angle $\theta$ . The
choice of functions
depends on whether $Y$ is compact or not. Correspondingly, in the
compact case, one can
choose the functions to be smooth $L^2(Y)$ with respect to the metric that
respects
the
circle
bundle, and to be smooth in the noncompact case. In the case $X$ and $Y$
are compact, one can decompose the eta function as a sum over
contribution
from a given $k$. For $k=0$, the phase is the same as the trivial circle
bundle case, $i^{h_{R.S.}^+}$, with the $+$ referring to positive
chirality,
and for $k\neq 0$ there is no contribution from $h$.

The contribution from $\eta$ is just the result of \cite{DMW1, DMW2},
which is
\(
\frac{\eta(s)}{2}=|R|^s \sum_{k=1}^{\infty}
\left( a k^{-(s-1)} + b k^{-(s-3)} + c k^{-(s-5)} \right)
\)

where the coefficients are given in terms of characteristic classes,
\begin{eqnarray}
a&=&c_1({\cal{L}}) \left(\rank(V(a)) \hat{A}_8 - \lambda^2 \right)
\\
b&=&\frac{2}{9} \lambda c_1^3(\cal{L})
\\
c&=&8 \frac{c_1^5(\cal{L})}{5!}
\end{eqnarray}

Then the above contributions combine as $\overline\eta_{R.S.}$ and can
be
inserted in
the phase \eqref{phase}.

\subsubsection{$E_8$ Dirac operator}

Now let us consider the $E_8$-coupled Dirac operator $D$ on $Y$.
Here we use the formalism
of Bismut-Cheeger \cite{BC} (and Dai \cite{Dai1}) for calculating the
adiabatic limit of the
reduced eta invariant.
Let ${\cal{R}}^X$ be the curvature of $X$ and $SX$ its spin bundle, with
spin connection $\nabla^X$ induced from the Levi-Civita connection on
$X$.
Associated to the principal $E_8$ bundle $P$ on $X$,
we have a  Hermitian vector bundle $V(a)$ as in section 3, with a
unitary
connection $\nabla^{V(a)}$. Then the bundle $SX \otimes V(a)$ has a
tensor product connection $\nabla=\nabla^X \otimes 1 + 1 \otimes
\nabla^{V(a)}$.
A natural representation
of
$Cl(X_p)$ on $SX_p$ can be extended to a representation on
$SX \otimes {V(a)}$.

Corresponding to the scaled metric $tg^X$ we have the Dirac operator
$D_{V(a)}^t$, whose reduced eta invariant, when
taken mod
$\ZZ$, was shown
by Bismut and Cheeger, and also by Dai, to be independent of $t$ and has
value
of
half the
index of $D_{V(a)}$. When ${\rm Ker}D_{Y/X}$ is a vector bundle on $X$,
one can use it to twist $D_X$. The connection on ${\rm Ker}D_{Y/X}$ is
obtained as the projection of a unitary connection on the
infinite-dimensional bundle
${\cal{E}}=L^2\left(\pi^{-1}(x),S_{\pi^{-1}(x)}\right)$
of smooth spinor sections along the $S^1$
fiber.
For $x\in X$, have $E_p^x=\iota_x^* E_p \rightarrow \pi^{-1}(x) \cong
S^1$
with $\pi^{-1}(x)
\buildrel \iota_x\over\hookrightarrow Y$. This assumption
that ${\rm Ker}D_{Y/X}$ is a vector bundle on $X$ implies that there is no spectral flow for
the family of Dirac operators on the fibers $D_{Y/X}$. This means that there are no anomalies
in this situation.
In this case, the adiabatic limit of the eta invariant on $Y$
has a closed formula given by
\footnote{Bismut and Cheeger \cite{BC} assumed that the family of 
Dirac operators $D_{Y/X}$ was invertible, but Dai
\cite {Dai1} just assumed that $\ker D_{Y/X}$ has constant rank,
which is what is stated here.}
${}^{,}$
\footnote{$\hat\eta$ that appears in this formula is renormalized by a
factor of $(\frac{1}{2\pi i})^{[\frac{p+1}{2}]}$ for a p-form.}

\(
\lim_{t\to \infty}\overline\eta(D_{V(a)}^t)
=\int_X \hat{A}\left( {\cal{R}}^X \right)
\wedge {\hat{\eta}}_{{V(a)}}  + \overline\eta (D_X \otimes {\rm ker}
D_{Y/X}) + \frac{1}{2}h'
\)
where 
$\hat{A}( {\cal{R}}^{X}) $ is the $\hat{A}$ invariant polynomial applied to
the curvature,
$\frac{1}{2} h'$ is a spin cobordism invariant, and
where the eta-form is a differential form on $X$ given by
\footnote{This is analogous to the Heat Kernel representation of the
eta invariant of $D$,
\(
\eta(D)=\frac{1}{\sqrt{\pi}}\int_{0}^{\infty}
{\rm tr} \left(D e^{-uD^2} \right) \frac{du}{u^{\frac{1}{2}}}
\nonumber \)
}

\(
\hat{\eta}_{{V(a)}} = \frac{1}{\sqrt{\pi}}
\int_{0}^{\infty} {\rm tr}^{even}
\left[ \left( D_{Y/X}+ \frac{c(T)}{4u} \right)
e^{-\left(B_u \right)^2} \right] \frac{du}{2 u^{\frac{1}{2}}} .
\)
Here $B_u:=\nabla^{V(a)}+ u^{\frac{1}{2}}D_{Y/X} -
\frac{c(T)}{4u^{\frac{1}{2}}}$ is the Bismut
superconnection (see e.g. \cite{BGV}), where $c$ denotes Clifford
multiplication and $T$ is the torsion of the connection.
The eta-form is of even degree, can be composed into homogenous even
parts as
$\hat{\eta}=\sum_{k=0}^{{\rm dim}X} \frac{1}{(2\pi i)^k}
[\hat{\eta}]_{2k}$
and has as the $0$-form component the $\eta$-invariant of the Dirac
operator
along the fiber.
This form arises as the
spectral correction to the families version of the Atiyah-Patodi-Singer
non-local elliptic boundary value problem. More precisely,
recall that $Z$ is a disk bundle over $X$, whose boundary is the
circle bundle $Y$ over $X$. The Bismut-Cheeger, Dai theorem in this
context asserts that
\(
{ch} \left({\rm Ind} ( D_{Z/X}) \right) =  \int_{\mathbb{D}}
\hat{A}({\cal{R}}^{Z/X}) \wedge {\hat{\eta}}_{V(a)} + 
({\rm boundary \; correction})
\in
H^{even}(X,
\mathbb  R) 
\label{ind}
\)
Here $\mathbb{D}$ is the disk which is the fiber of $Z$, $D_{Z/X}$ is
the family of
twisted Dirac operators along the fibers of $Z$
that are parametrized
by $X$ with the Atiyah-Patodi-Singer boundary conditions and
${\cal{R}}^{Z/X}$ is the curvature of the vertical tangent bundle of $Z$.
The last term in \eqref{ind} is a boundary correction term due to 
noninvertibility of the boundary operator.

The differential
\(
d\hat{\eta}=\int_{S^1} \hat{A}\left({\cal{R}}^{Y/X} \right)
\wedge {\rm tr}\left( e^{-\frac{1}{2\pi i} {\cal{R}}^{V(a)} }
\right)
\)
is closed (not exact) and  represents the odd Chern class of
$D_{Y/X}$. Here
${\cal{R}}^{Y/X}$ is the curvature of the connection on the vertical
tangent
bundle of $Y$,
and ${\cal{R}}^{V(a)}$ is the curvature of the unitary connection on
$V(a)$. After integrating over the fiber, we get an odd degree
differential form on $X$. This formula in particular implies that the
higher spectral flow vanishes.
\footnote{If $Y$ has a nonempty
boundary then \cite{Dai2} the results of Bismut and Cheeger still hold
provided one
keeps the invertibility condition.}

Even though we do not evaluate the above expression for the adiabatic
limit, we point that
we have reduced the adiabatic limit to an integral over the base, thus
relating
the M-theory data on the nontrivial circle bundle to the data of type
IIA
on $X$.


\section{Type IIA  theory}\label{IIA}
\subsection{The partition function}

Dimensional reduction of the eleven-dimensional action $I_{11}$ on
$S^1$ \cite{GP, CW, HN}, with a radius $R$, leads to
\footnote{the $S_{coupling}$ is the supersymmetric completion of
the action by algebraic terms in various fields, e.g. bilinear
and quartic in the fermions as well as coupling of the bilinear terms
to the p-forms.}
\footnote{Up to total derivative one can rewrite the Chern-Simons term
in terms of $H_3$ rather than $B_2$, namely $\int_X H_3 \wedge C_3
\wedge F_4$.}
\begin{eqnarray}
S_{IIA}&=&S_{NS}+S_{RR}+S_{C.S.} + S_{fermi} +S_{coupling}\\
S_{NS}&=&\frac{1}{2 \kappa_{10}^2} \int_X e^{- 2\phi}
\left[{\cal{R}} + 4d\phi \wedge *d\phi- \frac{1}{3}|H_3|^2 \right] 
\;d {\rm vol} \\
S_{RR}&=&\frac{1}{4 \kappa_{10}^2} \int_X 
\left[|F_2|^2 + |{{F}}_4|^2\right] \;d {\rm vol}\\
S_{C.S.}&=&-\frac{1}{4 \kappa_{10}^2} \int_X B_2 \wedge F_4 \wedge F_4 \\
S_{fermi}&=&-\frac{i}{2}\int_X \left[\bar{\psi}D_{R.S.}\psi +
\bar{\lambda}D\lambda\right] \;d {\rm vol}
\end{eqnarray}
where ${\cal{R}}$ is the scalar curvature of $X$, ${{F}}_4=G_4
- {\cal{A}} \wedge H_3$ is the gauge-invariant RR 4-form field strength, and
$D$ is the Dirac operator acting on the dilatino
$\lambda$, the superpartner of the dilaton $\phi$. The gravitational
coupling constant in eleven dimensions, $\kappa_{11}$, is related 
\footnote{explicitly $\kappa_{11}^2=2\pi R \kappa_{10}^2$ and $R=g_s
\sqrt{\alpha'}$ and $\kappa_{11}^2=\frac{1}{2} (2 \pi)^8 g_s^3
{\alpha'}^{\frac{9}{2}}$ give $\kappa_{10}^2=\frac{1}{2} (2 \pi)^7 g_s^2   
{\alpha'}^4$.}
to the one in ten dimensions by 
$\kappa_{10}^2=\frac{\kappa_{11}^2}{2 \pi R}$. As in the case for
M-theory, $S_{coupling}$ involves coupling of the fermions to the 
forms, as well as self-couplings, and we will not use this in this 
paper.

The partition function of type IIA string theory is of the form
\(
Z_{IIA} \sim Z_{NS} Z_{RR} Z_{C.S.} Z_{fermi} Z_{coupling}
\)

The Ramond-Ramond part is encoded in the theta
function $\Theta_{IIA}$ coming from summing over the
RR forms \cite{DMW1, DMW2}.
In \cite{MS} $Z_{NS}$ for flat potentials
    namely
\(
\exp\left[\pi i \int_X B_2^{(0)} F_4 F_4\right]
\)
as well as
$Z_{fermi}$, together with 1-loop determinants were considered. In
addition, the authors also include contribution to the partition 
function
from 1-loop corrections $\int_X B_2^{(0)} X_8$ to the effective
action
and the effect of membrane instantons. Here we focus on the
Ramond-Ramond
part and study the generalization to the case $[H_3] \neq 0$.

\subsection{Twisted $K$-theory}\label{loop} 

Our goal in this section is summarized in the following diagram.
\begin{equation} \label{eqAc2}
\xymatrix @=2pc { & E_8 \; \text{bundle over}\; Y + \text{(Bianchi)}\;
     \ar[dl]^{}
\ar[dr]_{} & \\
F \in K^0(X, H)   &   &    LE_8 \; \text{bundle over}\; X }
\end{equation}
This is the analog of what was described in section \ref{twistform}.

It has been suggested in \cite{Allan} that the $E_8$ bundle in M-theory
can be related to an $LE_8$ bundle in type IIA (on $X$).
Starting from  principal $E_8$ bundle over $Y$,
the dimensional reduction of the M-theory
     to type IIA gives a $LE_8$ bundle $P\p$ in ten dimenions,
characterized by the 3-form $H_3=\int_{S^1}G_4$ (or 
equivalently $H_3= \iota_v G_4$).

\(
\left\{
\begin{matrix}
E_8&\to & P\cr
&&\dwn\cr
S^1 & \to & Y\cr
&&\dwn\cr
& & X\cr
\end{matrix}
\right\}
\hsp{.2}\longrightarrow\hsp{.2}
\left\{
\begin{matrix}
LE_8&\to & Q\cr
&&\dwn\cr
& & X\cr
\end{matrix}
\right\}
\)
Note that since $E_8$ is an approximate $K(\mathbb Z, 3)$ up to
dimension 14,
it follows that principal $E_8$ bundles over $Y$ are classified by
$H^4(Y, \Z)$.
More precisely,
the characteristic class of the $E_8$ bundle is the restriction of the
first Pontjagin class $p_1$ to the 4-spheres in the 4-skeleton of
the base manifold,  $G_4 = \lambda(p_1) \in H^4(Y, \Z)$.
Then the class on $LE_8$ is $\pi_{*} \lambda(p_1) \in H^3(X, \Z)$.
There exists a $LE_8$ bundle, unique up to isomorphism, such that
the Dixmier-Douady class $DD(LE_8)=\pi_{*} \lambda(p_1)$. For $m \in
X$, $\pi^{-1}(m)=S^1$, one has
\(
C^{\infty}\left( \pi^{-1}(m),P|_{\pi^{-1}(m)}\right)\cong LE_8
\)

This gives the fibration above with
$
Q=\bigcup_{m \in X} C^{\infty}\left(
\pi^{-1}(m),P|_{\pi^{-1}(m)}\right)
$
so $DD(Q)= \pi_* \lambda(p_1) = H_3$.
The obstruction to lifting the $LE_8$ bundle $Q$ to an $\widehat{LE_8}$
bundle $P'$, covering $Q$, is the Dixmier-Douady class.
\(
\begin{matrix}
{\widehat{LE_8}}&\to & P' \cr
&&\dwn\cr
& & X\cr
\end{matrix}
\)
That is, such a  lift is possible
only when $H_3=dB_2$.
Therefore, in the presence of $F_0$, only the trivial case (in the sense
of the NS 3-form) can be seen in loop group picture.

We have seen in section \ref{twistform} that we can derive from the 4-form
$G_4$ on $Y$, a 4-form $F_4$ on $X$. Recalling that $F_2 = d\mathcal A$ and considering the inhomogeneous
even degree form\footnote{ since $F_0=0$ then there is no $F_{10}$.} 
$F = F_2 + F_4 + F_6 +F_8$, 
where $F_8= *_{10}F_2$ and $F_6 $ is obtained by dimensional reduction  of $G_7
= *G_4$
as in section \ref{twistform}, we have seen that $F$ is $d-H$ closed.
Then using the fact that
the twisted Chern character $ch_H : K^0(X, H) \to H^{even}(X, H)$ is an
isomorphism over the reals, we obtain an element $F \in K^0(X, H)
\otimes\mathbb R$.
Unfortunately, we do not know at this time how to lift this to a class
in $K^0(X, H)$,
as methods used when $H_3=0$ do not seem to apply in the twisted case.
We leave
this as an open problem.

\subsection{Twisted $K$-theory torus and theta functions}

In the presence of branes and and and $H$-flux, the RR fields
${{F}}$ are determined by
the twisted K-theory classes $x \in K(X, H)$ via the twisted Chern map
\cite{MM,Wi1,MW,BCMMS,Stev}
\(\label{F1}
\frac{{F}(x)}{2 \pi} = ch_{H}(x) \sqrt{{\hat A}(X)} \in
H^\bullet(X, H)
\)
where $\widehat{A}$ is the A-roof genus.

It turns out that the conjugate of $x$, $\bar x \in K(X, -H)$  
\(\label{F2}
\frac{{{F}(\bar x)}}{2 \pi} = ch_{-H}(\bar x) \sqrt{{\hat A}(X)}
\in H^\bullet(X, -H)
\)

Setting ${F} = \sum_{n=1}^{4} {{F}}_{2n}$ for the
gauge-invariant field strengths,
the RR field EOM  can be written succinctly as
\footnote{ In order to make the RR field strengths
homogeneous of
degree zero, one can \cite{F} use  K-theory with coefficients in
$K(pt)\otimes \R \cong \R\left[ [u, u^{-1}]\right]$ where the inverse
Bott element $u\in K^2(pt)$ has degree $2$, and look at the
corresponding chern character as a homomorphism of $\Z$-graded rings,
\(
ch : K(X,H) \rightarrow H^{even}\left(X,H;\R\left[ [u, u^{-1}]\right]
\right).
\nonumber \)
Then the total RR field strength is written as
\(
F= F_0 + u^{-1}F_2 +  u^{-2}F_4 + u^{-3}F_6 + u^{-4}F_{8}+ u^{-5}F_{10}.
\nonumber \)}
\(
d{{F}}=H_3 \wedge {F}
\)
Putting it another way,
the RR field EOM on the level of differential forms says that
the RR fields determine
elements in twisted cohomology,   $H^\bullet(X, H)$.
At the level of cohomology this implies $H_3 \wedge {{F}}_{n}=0$.
In $K_H$, or more precisely in the Atiyah-Hirzebruch spectral sequence
($AHSS$), this becomes
\footnote{There has also been proposals for S-duality-covariant
extensions of AHSS in \cite{uday, J1}.}
\(\label{square}
(H + Sq^3) \cup {{F}}_n=0
\)
\cite{DMW2} argue (and conjecture) that the M-theory partition function
on a circle bundle can be written in terms of fields satisfying
(\ref{square}).

A special case of the cup product pairing in twisted K-theory
followed by the standard index pairing of elements of K-theory
with the Dirac operator, explains the upper horizontal arrows
in the diagram,
\begin{equation} \label{rootA}
\begin{CD}
K^\bullet(X, H)  \times K^\bullet(X, -H)  @>>> K^0(X)
@>{\rm index}>>\ZZ \\
          @V{ch_H}\times {ch_{-H}}VV          @VV{ch} V      @VV{||}V \\
H^\bullet(X, H)  \times H^\bullet(X, -H)    @> >>
H^{even} (X)   @>\int_X{\widehat{A}(X)}\wedge>>\ZZ
\end{CD}\end{equation}
The bottom
horizontal arrows are cup product in twisted cohomology
followed by cup
product by ${\widehat{A}(X)} $
and by integration. By the Atiyah-Singer index theorem, the diagram
\eqref{rootA}
commutes.  Therefore the normalization given to the Chern character in
the definition of
$\frac{{{F}(x)}}{2 \pi} $
makes the pairings in twisted K-theory and
twisted cohomology isometric.

As noted by Witten \cite{Wi4}, there is a subtlety in the self-duality
$*{{F}}={{F}}$. It is in fact not possible to impose a
classical quantization
law on the periods of a self-dual p-form. This is because one cannot
simultaneously measure anticommuting periods, i.e. ones whose
intersection
number is non-zero.
The way around this is \cite{Wi4, MW} to interpret this self-duality as 
a
statement in the quantum theory and sum over half the fluxes, i.e. over
a maximal set of commuting periods. So we need
a phase space (in twisted K-theory) with a polarization or Lagrangian
subspace that naturally splits the
forms in half.
The lattice is $\Gamma_{K_H}=K(X,H)/K(X,H)_{tors}$. This is isomorphic
to the image of
the modified chern character homomorphism of ${\Z}_2$-graded rings,

\(
\sqrt{\hat{A}(X)} \wedge ch_H : K(X,H) \rightarrow H^{even}(X,H;\R)
\)
and the kernel is $K(X,H)_{tors}$, the torsion subgroup.
The lattice is unimodular by Poincar\'e duality in twisted K-theory.
In what follows, we give an analog of some of the constructions given
in \cite{DMW1} and \cite{MS}.

First, using equations (\ref{F1}) and (\ref{F2}), we get the metric (that
gives the kinetic energy)
\(
{{g}}(x,y)= \frac{1}{{2 \pi}^2} \int_X {F}(x) \wedge
*{F}(y)
\) 
which is defined on the lattice $\Gamma_{K_H}$, and which determines
a homogeneous metric on the twisted K-theory torus
${\mathcal T}_H(X) = (K(X,H) \otimes \mathbb R)/\Gamma_{K_H}$.

Similarly consider the bilinear form on the lattice $\Gamma_{K_H}$
\(
\omega(x,y)=\frac{1}{{2 \pi}^2} \int_X {F}(x) \wedge
{{F}}(\bar{y})=I(x \otimes {\bar{y}}), \qquad \forall\, x, y \in  K(X,H)
\)
where we notice that $x \otimes \bar y \in K(X)$. Here
\(
I(\xi)=\int_X \hat{A}(X) \wedge ch(\xi), \qquad \forall\, \xi \in K(X).
\)
For a torsion class $x_0 \in K(X, H)_{tors}$ and for any $x \in K(X, H)$, 
have $\omega(x,x_0)=0 = g(x, x_0)$ since
$nx_0=0$. This implies that $\omega(~,~)$ and $g(~,~)$ are well-defined
on the lattice $\Gamma_{K_H} = K(X, H)/K(X, H)_{tors} $.
If $X$ is a spin manifold and $\dim(X)=4n+2$, 
then $\omega$ is  antisymmetric: this essentially uses 
the arguments of \cite{Wi4} for the untwisted case.  
The only terms in $ch(\xi)$ that contribute to the value of $I(\xi)$ are
terms of degree $4k+2$ where $k\le n$, since $\hat{A}(X)$ has components only of degree
$4l$ for some $l$. That is, the only terms in $ch(\xi)$ that contribute to the value of $I(\xi)$ are
$ch_j(\xi)$ for $j$ odd. These terms are odd under the 
transformation $\xi \rightarrow \bar{\xi}$,
since $ch_j(\bar{\xi})=-ch_j(\xi)$ for $j$ odd, so
that $I(\xi)=-I(\bar{\xi})$. This implies $\omega(x,y)=I(x\otimes \bar{y})=
-I(y\otimes \bar{x})=-\omega(y,x)$ is antisymmetric. 
Then $\omega$ determines
a homogeneous differential 2-form on the twisted K-theory torus
${\mathcal T}_H(X)$ which is closed and integral.

The form $\omega$ is unimodular (i.e. $\frac{1}{2\pi i}\omega$ is integral and 
$\int_{\mathcal T_H(X)} e^{\frac{1}{2\pi i}\omega}=1$) on the lattice $\Gamma_{K_H}$
because $X$ is spin and because of Poincar\'e duality in twisted K-theory \cite{Ros},
i.e. the top line in (\ref{rootA}) is a unimodular pairing, and the lattice
$\Gamma_{K_H}$ is symplectic.  

The pair $(g, \omega)$ determine a K\"ahler form that is an
integral form. By the Kodaira embedding theorem 
\footnote{ which says that a compact complex manifold which admits a
positive line bundle can be holomorphically embedded in complex
projective space.},  the twisted K-theory torus $\mathcal T_H(X)$ is a smooth projective
algebraic variety.
Since the lattice $\Gamma_{K_H}$ is symplectic, 
it has a Lagrangian decomposition $\Gamma_1 \oplus \Gamma_2$ (or polarization)
in terms of commutative sublattices $\Gamma_1$ and $\Gamma_2$. 
Because of the duality between 
$\Gamma_1$ and $\Gamma_2$, there is an element
 $\theta_K \in \Gamma_1$ as argued in \cite{DMW1}, \cite{MS} so that, for any $y \in
\Gamma_2$ , $\Omega(y)=(-1)^{(\theta_K,y)}$. 
Summing over half the fluxes now amounts to summing over the
Lagrangian sublattice $\Gamma_1$, which gives
the type IIA partition function,
\begin{equation}\label{eqn:theta-omega}
\Theta_{IIA}(H: \tilde{\tau})=
e^{iu} 
\sum_{x \in \Gamma_1}
e^{i \pi {\tilde{\tau}_K}(x + \frac{1}{2} \theta_K)} \Omega(x)
\end{equation}
where $\tilde{\tau}$ are the period matrices when $F(x)$ is
replaced by ${F}(x)$. This is a theta
function on the torus $\mathcal T_H(X)$.
Explicitly, the quadratic
form on $\Gamma_1 \otimes \R$ is determined by:

\begin{eqnarray}
\Re{\tilde{\tau}}_K (x+ \frac{1}{2}\theta_K) &=& \frac{1}{(2 \pi)^2}
\int_X \left(-  {F}_2  {F}_8 +
 {F}_4 {F}_6\right)\\
\Im{\tilde{\tau}}_K (x+ \frac{1}{2}\theta_K) &=& \frac{1}{(2 \pi)^2}
\int_X \left( {F}_2\wedge *
{F}_2 + {F}_4\wedge *  {F}_4\right)
\end{eqnarray}
where $\Im\tilde \tau_K>0$, and
$u$ is given by
\(
u=-\pi \Re{\tilde{\tau}}_K( \frac{1}{2}\theta_K) 
\)
As in \cite{DMW1, MS}, the function $\Omega(x)$ given in equation \eqref{eqn:theta-omega} satisfies
the identity
\(
     \Omega(x + y) =  \Omega(x) \Omega(y) (-1)^{\omega(x,y)}
\)
There are potentially many such functions, but we will make a particular
choice as suggested in \cite{Wi4}. Let $q(V)$ denote the parity of the (real) dimension
of the space of chiral zero
modes of the real Dirac operator coupled to the real vector bundle $V$, $D_V$. It is a
topological invariant in $8n+2$ dimensions, and in particular for $X$.
Define $\Omega(x)=(-1)^{q(x \otimes \bar{x})}$, where we observe that
$x \otimes \bar{x} \in KO(X)$ for all $x\in K(X,H)$. Assuming $\Omega$ to be
identically one when restricted to $K(X, H)_{tors}$, it can then be
regarded
as a function on the lattice $ \Gamma_{K_H}$. By the argument in \cite{Wi4}, it 
determines a holomorphic and hermitian line
bundle $\cL$ over ${\cT}_H (X)$, with a connection having curvature
equal to $\omega$. $\cL$ has a holomorphic section $\Theta$ as defined above
which is unique (up to multiplication by scalars) 
by the Riemann-Roch theorem which says in this case that
$\dim H^0({\cT}_H (X), \cL) = \int_{{\cT}_H (X)} e^{c_1(\cL)} = 1$, 
since $c_1(\cL) = \frac{1}{2\pi i} \omega$ and $\omega$ is unimodular.
Notice that  if we changed the spin structure on $X$, then the
twisted K-theory torus $\mathcal T_H(X)$ doesn't change, but what changes is the
choice of section of the line bundle $\cL$ over $\mathcal T_H(X)$, i.e. the
theta function changes, since $\Omega$ depends on the choice of spin structure. 
In the case when $X = W \times \Sigma$
where $W$ is a compact spin 8 dimensional manifold and $\Sigma$ is a compact Riemann surface, 
then the (mod 2) index is often a nontrivial function on the subspace of spin structures arising
from $\Sigma$.

For the argument above to work, we need to know that the function $\Omega$ is constant on the
torsion subgroup $K({X}, H)_{tors}$. This may not be satisfied
in general, and so may give rise to an anomaly. However, in an
analogous setting, Hopkins and Singer \cite{HS} remove this anomaly by
a deeper analysis of the situation.


  To connect with the dimensional reduction metric ansatz, it is
convenient, as in \cite{DMW1, MS},
to choose a polarization that keeps only positive powers of $t$ in the
kinetic term upon scaling $g_X \rightarrow t g_X$ under which $\int_X
*1|F_{2p}|^2
\rightarrow
t^{(5-2p)} \int_X *1 |F_{2p}|^2$ . Since coefficients of $|F_{2p}|^2$
for $p\geq 3$ tend to zero in the limit $t \rightarrow \infty$, we keep
positive powers in
the expansion

\(
\sum_{p=1}^{4} t^{(5-2p)}|F_{2p}|^2
\)

The correction due to Lagrangian or polarization amounts to
shifting the class $x \in \Gamma_1$ by the  theta element $\frac{1}{2}\theta_K$,
\(
\left[\frac{{{F}}(x)}{2 \pi}\right]=ch_H (x + \frac{1}{2}\theta_K)
\sqrt{{\hat
A}(X)}.
\)

To get only positive powers in $t$, we take $\Gamma_1$ to be the
complementary lattice to the lattice $\Gamma_2$ that consists of
K-theory
classes $x$ such that $ch_n(x)=0$ for $n=0,1,2$. The dominant
contribution
comes from the K-theory class $x\in \Gamma_1$ with $F_2(x)=0$ (still
assuming $F_0=0$).
The
sublattice of such classes with virtual dimensions $0$
such that $c_1(x)=0$ is $\Gamma'_1$. Then the IIA partition function
reduces to

\(
\Theta_{IIA}(H:\tilde\tau)=e^{iu}
\sum_{x \in \Gamma'_1} e^{-\pi t \int_X ||{{F}}_4||^2} e^{i\pi
\int_X {{F}}_4  {{F}}_6}
\Omega(x)
\)

The corresponding modification to the phase $\Im(S_{IIA})= -2\pi
{\tilde\Phi}$ of the action (5.1) is found by expanding ${{F}}_4 {{F}}_6$
and picking the appropriate degree. We note that many of the computations
in \cite{MS} that were performed for the cohomologically trivial case,
extends to the nontrivial case as well. For example,
\(
{\tilde\Phi}=\Phi + \frac{1}{8 \pi^2} \left[B_2 F_4^2 + B_2 F_2 F_6
+ \frac{1}{3}B_2^3F_2^2 \right]
\)
where $\Phi$ is the result of \cite{DMW1}, still holds, eventhough it was 
derived when $H_3=dB_2$. Basically, this is due to the fact that by
lifting to
the total space of the $LE_8$ bundle, $H_3$ becomes $dB_2$, by the general 
property of characteristic classes (recall that $H_3$ is the
Dixmier-Douady 
characteristic class). 
One might be interested in setting $F_2=0$ to simplify the above
expression, in which case, the contribution is just the one coming
from the Chern-Simons part, $S_{C.S.}$, of the Type IIA action.


Consider a $PU$ bundle $E$ over $X$, with 3-curvature 
 $H$, cf. \cite{BCMMS}. Then any other  3-curvature $H'$
for $E$, has the property that $H' = H + dB$ where $B$ is a
global
2-form on $X$, so that $K(X,H)\cong K(X, H+dB).$ Also if 
$E'$ is another $PU$ bundle over $X$ with 3-curvature $H''$,
then again we have that $H'' = H + dB'$, where $B'$ is a global
2-form on $X$, so that $K(X,H)\cong K(X, H+dB')$, cf. \cite{BCMMS}.
In particular, this implies that the twisted $K$-theory tori are isomorphic,
\(
{\cT}_H (X)\cong {\cT}_{H+dB} (X).
\)

A gauge transformation for $B$, $B_2 \rightarrow B_2 + f_2$,
$f_2 \in H^2(X,\Z)$ will leave ${F}$ invariant if this gauge
transformation
also acts on $K(X,H)$ as $x \rightarrow \pi^*\cL (-f_2) \tensor \tilde{x}$,
$\tilde{x} \in K(X,H)$,
where the line bundle $\cL (-f_2)$ has a Chern class given by
$c_1 (\cL (-f_2))=-f_2$.
Then this gauge transformation acts as an automorphism of
$\Gamma_K$,
preserving the symplectic form $\omega$ on the twisted $K$-theory torus.

\section{Discussion}\label{dis}

  In this paper we considered relating the fields of M-theory to those
of Type IIA in the large volume limit, for a nontrivial circle bundle
and in the presence of nontrivial NS flux $H_3$. We derived the RR
fields
of Type IIA from the M-theory 4-form $G_4$ satisfying Maxwell-type
equations, and have shown that those fields are elements in twisted 
cohomology $H^{even}(X, H_3)$. In order to write the partition
function of Type IIA, we constructed the twisted K-theory torus.
We also have considered the topological 
part of the M-theory partition function, and, for the general case when
the $E_8$ vector bundle that twists the Dirac operator, is not lifted
from the
base, and the NS field $H_3$ is nontrivial in cohomology, 
we have written the eta invariant (that determines the phase) in the 
adiabatic limit as an integral in Type IIA. We have also discussed 
the similarities and differences with the cases $B=0$ \cite{DMW1} and
$H=dB$ \cite{MS}, considered before.

In relating the eta invariant in M-theory (i.e. on $Y$) to an integral in
type IIA (on $X$), we naturally encountered the eta-forms in the
integrand. It would be very interesting to compute the eta-forms for
nontrivial circle bundles $Y$ over $X$ for the case when the Dirac
operator on the total space is coupled
to
vector bundles that are not lifted from the base. Zhang \cite{Zhang}   
has computed $\hat\eta$ in the lifted case.
It was also computed by Goette \cite{Gt} using
G-equivariant eta invariants
for the nontrivial circle bundle. It might also be interesting to
give a physical interpretation to the various form components of
$\hat\eta$ that show up in the phase, perhaps in analogy to Witten's
global anomaly for the degree zero component.

T-duality \cite{Bush1, Bush2} relates Type IIA to type IIB string
theory and can be
implemented at the level of the effective action \cite{effective},
and K-theory \cite{Hori, BEM}.
The RR fields of type IIB, in
the presence of the NS field $H_3$ and in the
absence of branes, are determined by an element $\hat{x} \in
K_H^1(X)$.
Since $K_H^1(X)\cong {\widetilde{K}}_H^0(S^1 \times X)$, one
can view \cite{MW}
$ch_H(x)$ as an even class on $S^1 \times X$, which upon
integration
over $S^1$ gives an odd element $i_*ch_H(x)$.
This is what
one expects from $T$-duality for the case $X$ is the total space
of
a circle bundle. Then T-duality maps the 4-form $G_4$ of
type IIA to the self-dual 5-form
$F_5$ of type IIB. It is then resonable to believe that one can relate
properties of one to those of the other. In fact, as explained by Witten
\cite{Wi4}, it is possible to deduce the integrality conditions on $G_4$
from
the
quantum mechanics of the self-dual five-form field strength.
It would be interesting to implement T-duality of IIA/IIB at the
level of K-theory theta functions, and deduce the T-duality anomalies,
along the lines of \cite{MS}, from the IIB picture, or alternatively,
follow the methods of \cite{HS}.

We have constrained our discussion mostly to the RR part of the
partition function (twisted by $H_3$). It would be interesting to include
the other parts of the partition function, e.g. fermions, one-loop and 
quantum
corrections, and find the corresponding (T-duality) anomaly-free 
partition
function. In order to get a T-duality anomaly-free partition functions,
\cite{MS} 
considered super-theta functions by making the K-theory torus
into a supertorus. It would be interesting to describe such functions
in this context.

In trying to derive the answer from the $LE_8$ picture, one can give
a description of the twisting field $H_3$ and the RR 4-form field
that is being twisted, but only in twisted cohomology. It would be
interesting
to see how lifting problem might be solved in the twisted case, and give
the corresponding description in twisted K-theory (see subsection
\ref{loop}).

In order to be able to define the theta functions, one needed the
condition on the ${\ZZ}_2$-valued function $\Omega$ to be identically
one on $K(X,H)_{\rm torsion}$, in order to descend to a function on
$\Gamma_{K_H}$ and thus define $\cal{L}$ and the theta functions. 
In \cite{DMW1}, this condition was shown to be equivalent to a condition
on the Stiefel-Whitney class $W_7(X)=0$. If $\Omega\neq1$ then there is a 
possible anomaly.
It 
would
be
interesting to see, as suggested from the work of Hopkins and
Singer \cite{HS}, how a
refinement of the construction shows the absence of such an anomaly,
in the presence of the twisting by $H_3$.

The anomaly cancellation condition, in the untwisted case,
$Sq^3=0$, 
shows up as an obstruction to lifting cohomology to K-theory and also as a
condition for modding out by torsion in the M-theory phase (averaging
over the torsion classes) as in
\cite{DMW1, MS}. It would be interesting to see if the corresponding 
condition
in the twisted case  \cite{MMS1, MMS2, MW} $Sq^3 + [H]=0$ can be viewed as
an obstruction to lifting to twisted K-theory (beyond AHSS), and as
the condition for the M-theory phase, and also perhaps as an obstruction
to creating $LE_8$ bundles. Obviously, a lot of work needs to be done, and
we hope to address some of those interesting problems in the future.

\bigskip\bigskip
\noindent
{\bf Acknowledgements.}
\noindent
V.~M. thanks I.~M.~Singer for mathematical discussions related
to this paper, and H.~S. thanks I.~Kriz for useful earlier
discussions. We thank A.~Bergman and U.~Varadarajan for a useful 
question and comment on the first version of the paper. 
We thank the Australian Research Council for support.

\noindent


\end{document}